\documentclass[twocolumn,superscriptaddress,prc,showpacs,amsmath,amssymb,letterpaper]{revtex4-2}
\usepackage{graphicx,color}
\usepackage{amssymb}
\usepackage{xspace}
\usepackage{amsmath}
\usepackage{latexsym}
\usepackage{amsxtra}
\usepackage{amscd}
\usepackage{dcolumn}
\usepackage{bm}
\usepackage{threeparttable}
\usepackage{dcolumn}

\newcolumntype{d}[1]{D{.}{.}{#1}}
\newcolumntype{k}{D{(}{(}{-3}}
\usepackage[breaklinks]{hyperref}
\usepackage{url}
\usepackage{breakurl}
\graphicspath{{./}}
\let\oldsqrt\sqrt
\def\sqrt{\mathpalette\DHLhksqrt}
\def\DHLhksqrt#1#2{\setbox0=\hbox{$#1\oldsqrt{#2\,}$}\dimen0=\ht0
\advance\dimen0-0.2\ht0
\setbox2=\hbox{\vrule height\ht0 depth -\dimen0}%
{\box0\lower0.4pt\box2}}

\newcommand{\nuc}[2]{$^{#1}$#2}

\definecolor{nicegreen}{RGB}{46,204,64}

\begin{document}
\title{Single-particle strength toward \texorpdfstring{$\boldsymbol{N=32}$}{$N=32$}: \\
  Spectroscopy of \texorpdfstring{$\boldsymbol{^{51}}$}{${^{51}}$}Ca via the \texorpdfstring{$\boldsymbol{^{50}}$}{${^{50}}$}Ca\texorpdfstring{$\boldsymbol{(d,p)}$}{(d,p)} reaction}
\author{C. Ferrera}
\affiliation{Instituto de Estructura de la Materia, CSIC, E-28006 Madrid, Spain}%
\author{K.~Wimmer}\email{k.wimmer@gsi.de}
\affiliation{GSI Helmholtzzentrum f\"ur Schwerionenforschung GmbH, Planckstr. 1, D-64291 Darmstadt, Germany}%
\affiliation{Department of Physics, The University of Tokyo, Hongo, Bunkyo-ku, Tokyo 113-0033, Japan}
\author{D.~Suzuki}
\altaffiliation{Department of Physics and Quark Nuclear Science Institute, the University of Tokyo, 7-3-1 Hongo, Bunkyo, Tokyo, 113-0033, Japan}
\affiliation{RIKEN Nishina Center, 2-1 Hirosawa, Wako, Saitama 351-0198, Japan}%
\author{N.~Imai}
\affiliation{Center for Nuclear Study, University of Tokyo, 2-1 Hirosawa, Wako-shi, 351-0106, Saitama, Japan}%
\author{A.~Jungclaus}
\affiliation{Instituto de Estructura de la Materia, CSIC, E-28006 Madrid, Spain}%
\author{T.~Miyagi}
\affiliation{Center for Computational Sciences, University of Tsukuba, 1-1-1 Tennodai, Tsukuba 305-8577, Japan}%
\author{Y.~Utsuno}
\affiliation{Advanced Science Research Center, Japan Atomic Energy Agency, Tokai, Ibaraki, 319-1195, Japan}%
\author{D.~Das}
\affiliation{Institut f\"ur Kernphysik, Technische Universit\"at Darmstadt, 64289 Darmstadt, Germany}
\affiliation{GSI Helmholtzzentrum f\"ur Schwerionenforschung GmbH, Planckstr. 1, D-64291 Darmstadt, Germany}%
\author{T.~Chillery}
\affiliation{Center for Nuclear Study, University of Tokyo, 2-1 Hirosawa, Wako-shi, 351-0106, Saitama, Japan}%
\author{S.~Hanai}
\altaffiliation{Department of Physics, Institute of Science Tokyo, Tokyo 152-8550, Japan}
\affiliation{Center for Nuclear Study, University of Tokyo, 2-1 Hirosawa, Wako-shi, 351-0106, Saitama, Japan}%
\author{J.W.~Hwang}
\affiliation{Center for Exotic Nuclear Studies, Institute for Basic Science, 55, Expo-ro, Yuseong-gu, Daejeon, 34126, Republic of Korea}%
\author{N.~Kitamura}
\affiliation{Center for Nuclear Study, University of Tokyo, 2-1 Hirosawa, Wako-shi, 351-0106, Saitama, Japan}%
\author{R.~Kojima}
\affiliation{Center for Nuclear Study, University of Tokyo, 2-1 Hirosawa, Wako-shi, 351-0106, Saitama, Japan}%
\author{S.~Michimasa}
\author{R.~Yokoyama}
\affiliation{Center for Nuclear Study, University of Tokyo, 2-1 Hirosawa, Wako-shi, 351-0106, Saitama, Japan}%
\author{Y.~Anuar}
\affiliation{Center for Nuclear Study, University of Tokyo, 2-1 Hirosawa, Wako-shi, 351-0106, Saitama, Japan}%
\author{M.~Armstrong}
\affiliation{ Institut f\"ur Kernphysik, Universit\"at zu K\"oln, Cologne 50937, Germany}%
\affiliation{GSI Helmholtzzentrum f\"ur Schwerionenforschung GmbH, Planckstr. 1, D-64291 Darmstadt, Germany}%
\author{S.~Bae}
\affiliation{Center for Nuclear Study, University of Tokyo, 2-1 Hirosawa, Wako-shi, 351-0106, Saitama, Japan}%
\author{Y.~Cho}
\affiliation{Center for Exotic Nuclear Studies, Institute for Basic Science, 55, Expo-ro, Yuseong-gu, Daejeon, 34126, Republic of Korea}%
\author{M.~Dozono}
\affiliation{RIKEN Nishina Center, 2-1 Hirosawa, Wako, Saitama 351-0198, Japan}%
\affiliation{Graduate School of Science, Kyoto University, Yoshida-honmachi, Sakyo, Kyoto 606-8501, Japan}%
\author{F.~Endo}
\affiliation{Research Center for Nuclear Physics, Osaka University, Ibaraki, 567-0047 Osaka, Japan}%
\affiliation{RIKEN Nishina Center, 2-1 Hirosawa, Wako, Saitama 351-0198, Japan}%
\author{S.~Escrig}
\affiliation{Instituto de Estructura de la Materia, CSIC, E-28006 Madrid, Spain}%
\author{N.~Fukuda}
\affiliation{RIKEN Nishina Center, 2-1 Hirosawa, Wako, Saitama 351-0198, Japan}%
\author{T.~Haginouchi}
\affiliation{Department of Physics, Tohoku University,  6-3 Aramaki-aza-Aoba, Aoba-ku, Sendai, Miyagi 980-8578, Japan}
\author{S.~Hayakawa}
\affiliation{Center for Nuclear Study, University of Tokyo, 2-1 Hirosawa, Wako-shi, 351-0106, Saitama, Japan}%
\author{Y.~Hijikata}
\affiliation{RIKEN Nishina Center, 2-1 Hirosawa, Wako, Saitama 351-0198, Japan}%
\author{G.~Ikemizu}
\affiliation{Graduate School of Science, Kyoto University, Yoshida-honmachi, Sakyo, Kyoto 606-8501, Japan}%
\author{S.~Ishio}
\affiliation{Department of Physics, Tohoku University,  6-3 Aramaki-aza-Aoba, Aoba-ku, Sendai, Miyagi 980-8578, Japan}
\author{A.~Kasagi}
\affiliation{Graduate School of Artificial Intelligence and Science, Rikkyo University, 3-34-1 Nishi Ikebukuro, Toshima-ku, Tokyo 171-8501, Japan}%
\affiliation{High Energy Nuclear Physics Laboratory, Cluster for Pioneering Research, RIKEN, 2-1 Hirosawa, Wako, Saitama 351-0198, Japan}%
\author{K.~Kawata}
\affiliation{Center for Nuclear Study, University of Tokyo, 2-1 Hirosawa, Wako-shi, 351-0106, Saitama, Japan}%
\author{J.~Li}
\affiliation{Center for Nuclear Study, University of Tokyo, 2-1 Hirosawa, Wako-shi, 351-0106, Saitama, Japan}%
\author{S.~Masuoka}
\affiliation{Center for Nuclear Study, University of Tokyo, 2-1 Hirosawa, Wako-shi, 351-0106, Saitama, Japan}%
\author{B.~Moon}
\affiliation{Center for Exotic Nuclear Studies, Institute for Basic Science, 55, Expo-ro, Yuseong-gu, Daejeon, 34126, Republic of Korea}%
\author{K.~Okawa}
\affiliation{Center for Nuclear Study, University of Tokyo, 2-1 Hirosawa, Wako-shi, 351-0106, Saitama, Japan}%
\author{S.~Ota}
\affiliation{Center for Nuclear Study, University of Tokyo, 2-1 Hirosawa, Wako-shi, 351-0106, Saitama, Japan}%
\affiliation{Research Center for Nuclear Physics, Osaka University, Ibaraki, 567-0047 Osaka, Japan}%
\author{H.~Qin}
\affiliation{Center for Nuclear Study, University of Tokyo, 2-1 Hirosawa, Wako-shi, 351-0106, Saitama, Japan}%
\author{T.~Saito}
\affiliation{Center for Nuclear Study, University of Tokyo, 2-1 Hirosawa, Wako-shi, 351-0106, Saitama, Japan}%
\author{A.~Sakaue}
\affiliation{Center for Nuclear Study, University of Tokyo, 2-1 Hirosawa, Wako-shi, 351-0106, Saitama, Japan}%
\author{H.~Sakurai}
\affiliation{RIKEN Nishina Center, 2-1 Hirosawa, Wako, Saitama 351-0198, Japan}%
\author{Y.~Shimizu}
\affiliation{RIKEN Nishina Center, 2-1 Hirosawa, Wako, Saitama 351-0198, Japan}%
\author{S.~Shimoura}
\affiliation{Center for Nuclear Study, University of Tokyo, 2-1 Hirosawa, Wako-shi, 351-0106, Saitama, Japan}%
\affiliation{RIKEN Nishina Center, 2-1 Hirosawa, Wako, Saitama 351-0198, Japan}%
\author{Y.~Son}
\affiliation{Center for Exotic Nuclear Studies, Institute for Basic Science, 55, Expo-ro, Yuseong-gu, Daejeon, 34126, Republic of Korea}%
\author{T.~Sumikama}
\affiliation{RIKEN Nishina Center, 2-1 Hirosawa, Wako, Saitama 351-0198, Japan}%
\author{H.~Suzuki}
\affiliation{RIKEN Nishina Center, 2-1 Hirosawa, Wako, Saitama 351-0198, Japan}%
\author{H.~Takeda}
\affiliation{RIKEN Nishina Center, 2-1 Hirosawa, Wako, Saitama 351-0198, Japan}%
\author{Y.~Togano}
\affiliation{RIKEN Nishina Center, 2-1 Hirosawa, Wako, Saitama 351-0198, Japan}%
\author{J.~Vesic}
\affiliation{Jozef Stefan Institute, Ljubljana 1000, Slovenia}%
\author{K.~Yako}
\affiliation{Center for Nuclear Study, University of Tokyo, 2-1 Hirosawa, Wako-shi, 351-0106, Saitama, Japan}%
\author{Y.~Yamamoto}
\affiliation{Center for Nuclear Study, University of Tokyo, 2-1 Hirosawa, Wako-shi, 351-0106, Saitama, Japan}%
\author{K.~Yoshida}
\affiliation{RIKEN Nishina Center, 2-1 Hirosawa, Wako, Saitama 351-0198, Japan}%
\author{M.~Yoshimoto}
\affiliation{RIKEN Nishina Center, 2-1 Hirosawa, Wako, Saitama 351-0198, Japan}%

\begin{abstract}
States in the neutron-rich isotope \nuc{51}{Ca} were populated via the \nuc{50}{Ca}$(d,p)$ transfer reaction in inverse kinematics at a beam energy of about 14~$A$MeV. The experiment was performed using a decelerated radioactive \nuc{50}{Ca} beam from the OEDO facility and the TiNA2 silicon array in combination with the SHARAQ magnetic spectrometer at RIBF/RIKEN. The energies of excited states in \nuc{51}{Ca} were reconstructed via missing mass spectroscopy, and angular distributions of protons were measured to extract differential cross sections. From a comparison with adiabatic distorted wave approximation (ADWA) calculations, spectroscopic factors were deduced for several states, including the ground state and excited states up to 4.2~MeV. These results are compared with shell-model calculations, as well as ab initio valence-space in-medium similarity renormalization group (VS-IMSRG) predictions. The data support the assignment of the $1/2^-$ and $5/2^-$ single-particle states and provide evidence for a candidate $9/2^+$ state with a structure consistent with neutron excitation into the $0g_{9/2}$ orbital. These findings contribute new constraints on the single-particle structure and shell evolution in neutron-rich calcium isotopes.
\end{abstract}
\maketitle
\section{Introduction}

The neutron-rich calcium isotopes have become a central testing ground for modern nuclear structure theory, especially in the context of shell evolution and the emergence of new magic numbers far from stability. The drop in neutron separation energy observed through precision mass measurements~\cite{wienholtz13,michimasa18} for $N=32$ and 34 provided the first clear evidence for new shell closures, that are not observed in the valley of stability. This result is supported by $\gamma$-ray spectroscopy, which confirmed a high excitation energy of the first $2^+$ states in \nuc{52}{Ca}~\cite{huck85} and \nuc{54}{Ca}~\cite{steppenbeck13}.
The theoretical prediction of a new shell closure for neutron number 34 is based on the strongly attractive interaction between protons in the $0f_{7/2}$ and neutrons occupying the $0f_{5/2}$ orbital. Reduced occupation of the former in the Ca isotopes, compared to Ni and nuclei in the valley of stability, leads to an increase in energy of the $\nu 0f_{5/2}$ orbital giving rise to the shell closures at $N=32$ and 34~\cite{steppenbeck13}. This also explains why the validity of the new magic numbers is very local applying only to Ca, and to some extend to its immediate neighbors~\cite{liu19}.
The discovery of \nuc{60}{Ca} and the identification of a weakly bound ground state~\cite{tarasov18} extended the reach of experimental investigations toward $N=40$, and challenged theoretical models to describe the evolution of nuclear structure in this extremely neutron-rich regime.
These findings indicate the appearance of new magic numbers at $N=32$ and $N=34$, attributed to the repulsive nature of the tensor~\cite{otsuka01} and three-nucleon forces~\cite{holt12} between protons and neutrons occupying specific orbitals.  While these developments mark major successes, achieving a consistent description across the entire chain of calcium isotopes, from stability to the neutron dripline, remains an open challenge.
In particular, there is a significant lack of experimental data on the single-particle structure in the neutron-rich region.
Spectroscopic information on excited states, including spin-parity assignments and spectroscopic factors from direct reactions, is essential to benchmark and constrain theoretical models in this region.
The goal of the present work is to provide robust experimental data on the location and strength of single-particle states in neutron-rich Ca isotopes, starting with \nuc{51}{Ca} at neutron number $N=31$.

The first spectroscopy of excited states in \nuc{51}{Ca} was performed via particle spectroscopy following three-neutron transfer reactions with heavy-ion beams~\cite{catford88}. However, the low-lying states observed in that study could not be confirmed by later experiments.
The first assignment of $\gamma$ rays to excited states in \nuc{51}{Ca} came from studies of the $\beta$ decay of neutron-rich potassium isotopes~\cite{perrot06}. Although spin and parity assignments were not possible at the time, neutron$-\gamma$ and $\gamma-\gamma$ coincidence measurements enabled the construction of a level scheme.
This level scheme was subsequently confirmed and extended through multinucleon transfer reactions~\cite{rejmund07,fornal08}, which identified candidates for the single-particle $1/2^-$ and $5/2^-$ states at 1.718 and 3.478~MeV, respectively.
Shell-model calculations using the $fp$ model space and two different effective interactions indicated that these states have dominant $\nu(1p_{3/2})^2(1p_{1/2})^1$ and $\nu(1p_{3/2})^2(0f_{5/2})^1$ configurations, where the two neutrons in the $1p_{3/2}$ orbital are coupled to $J=0$~\cite{rejmund07}. Other excited states in \nuc{51}{Ca} are attributed to configurations in which the two neutrons are coupled to $J=2$ and are therefore not expected to be populated in the present transfer reaction.
A candidate for a positive-parity $9/2^+$ state was reported at 4.155~MeV~\cite{fornal08}, with its structure believed to be dominated by proton excitations from the $sd$ shell to the $0f_{7/2}$ orbital. In contrast, inelastic proton scattering measurements~\cite{riley16} identified the $9/2^+$ state as a member of a multiplet arising from the coupling of the odd neutron to the octupole $3^-_1$ excitation in the \nuc{50}{Ca} core.
Inverse-kinematics quasi-free $(p,2p)$ scattering reactions on a \nuc{52}{Ca} beam were used to probe neutron-hole states~\cite{enciu22}. These experiments revealed strong population of the $7/2^-$ state at 3.453~MeV, consistent with neutron removal from the core $0f_{7/2}$ orbital.
The ground-state spin and parity of \nuc{51}{Ca} were definitively determined to be $3/2^-$ via laser spectroscopy measurements including its magnetic and quadrupole moments~\cite{garciaruiz15}. This result provides a model-independent anchor for shell-model descriptions of neutron-rich calcium isotopes.

In this work, we address the single-particle character of states in \nuc{51}{Ca} by employing a direct one-neutron transfer reaction on \nuc{50}{Ca}. This direct transfer reaction provides sensitivity to the orbital angular momentum and spectroscopic strength of individual states, making it a powerful tool for probing their single-particle character. The aim of the present study is to confirm spin-parity assignments of previously suggested  $1/2^-$ and $5/2^-$ states, and to test the $9/2^+$ assignment of a positive-parity candidate state. The extracted differential cross sections and spectroscopic factors offer new insights into the occupancy of neutron orbitals and the evolution of shell structure in neutron-rich calcium isotopes.

\section{Experimental setup}
The experiment was conducted at the RI Beam Factory, operated by the RIKEN Nishina Center and the Center for Nuclear Study (CNS), The University of Tokyo. Due to issues with beam acceleration, the experiment was carried out in two separate campaigns in 2022 and 2024. While the overall setup and secondary beam properties were largely similar, key differences are summarized in Table~\ref{tab:campaigns} and described in detail in this section.
\begin{table}[h]
  \caption{Summary of the two experimental campaigns.} 
  \label{tab:campaigns}
  \centering
  \begin{tabular}{lrr}
    \hline
           & 2022 & 2024\\
    \hline
    \hline
    Secondary beam energy ($A$MeV) & 14.2 & 13.1\\
    Energy spread FWHM ($A$MeV) & 7.1 & 6.1\\
    OEDO transmission FE9-FE12 (\%) & 58.1 & 58.4\\  
    OEDO \nuc{50}{Ca} purity at S0 (\%) & 77.1 & 77.6 \\
    CD$_2$ target thickness ($\mu$g/cm$^2$) & 260 & 644 \\
    \nuc{50}{Ca} beam intensity on target ($10^3$ pps) & $16.9$ & $13.1$\\
    SHARAQ transmission S0-S1 (\%) & 14.7 & 35.8\\
    Run time (h) & 89.5 & 45.5\\
    \hline
  \end{tabular}
  \end{table}

  \subsection{Beam production and transport}
  The secondary beam was produced by projectile fragmentation of a 345~$A$MeV \nuc{70}{Zn} primary beam impinging on a 10-mm-thick Be target. The resulting cocktail beam was separated using the BigRIPS separator~\cite{kubo12} and transported to the Optimized Energy Degrading Optics (OEDO) beamline~\cite{michimasa19}. Identification of the incoming beam particles was achieved event-by-event by combining time-of-flight and position measurements. The time-of-flight was determined between a poly-crystalline chemical vapor deposition (CVD) diamond detector located at the F3 achromatic focus~\cite{michimasa13} and a pair of strip-readout parallel-plate avalanche counters (SR-PPACs) at the FE9 dispersive focal plane~\cite{hanai23}. The beam trajectory at FE9 was also measured using the same SR-PPACs. See Fig.~3 in Ref.~\cite{michimasa19} for the location of the focal planes.
  The beam consisted mainly of \nuc{50}{Ca}, with a purity of 69\% in 2022 and 73\% in 2024. The beam energy was then degraded from approximately 170~$A$MeV by inserting a wedge shaped degrader~\cite{hwang19} (central thickness 11.8 mm, angle 27~mrad) at FE9. The final kinetic energies of the \nuc{50}{Ca} beam particles were 14.2~$A$MeV (with a full width at half maximum [FWHM] energy spread of 7.1~$A$MeV) in 2022 and 13.1~$A$MeV (FWHM of 6.1~$A$MeV) in 2024. The radio-frequency deflector was not employed in the present study.
  The beam was subsequently transported through the OEDO beamline to the secondary target located at the S0 focus of the SHARAQ spectrometer~\cite{uesaka12}. The transmission between FE9 and FE12 was approximately 58\% and the purity of \nuc{50}{Ca} at S0 about 77\% in both experiments. CD$_2$ targets with a diameter of 50~mm were installed at S0, and to increase the luminosity, the target thickness was increased from 260~$\mu$g/cm$^{2}$ in 2022 to 644~$\mu$g/cm$^{2}$ in 2024. The average \nuc{50}{Ca} beam intensity amounted to 16.9 (13.1) kpps in 2022 (2024). The beam was tracked event-by-event by a pair of SR-PPACs located 1962~mm and 1462~mm upstream of the secondary target, respectively. These SR-PPACs also provided timing information for measuring the velocity of the beam particles after the degrader. The two tracking SR-PPACs at FE12 also provided the incident angle and trajectory of the beam ions on an event-by-event basis. 
  Two experimental triggers were employed for data acquisition. The F3 downscaled trigger fired for a fraction, fixed by the downscale factor, of the events in which a beam ion was detected at F3, allowing for beam intensity and transmission determination among other beam properties. The second trigger only fired when an ion was detected at F3 in coincidence with a charge deposition in TiNA2. The \nuc{50}{Ca}$(d,p)$\nuc{51}{Ca} reaction events of interest were extracted from this trigger. 

\subsection{TiNA2 charged particle array}
\begin{figure}[!h]
  \includegraphics[width=\columnwidth]{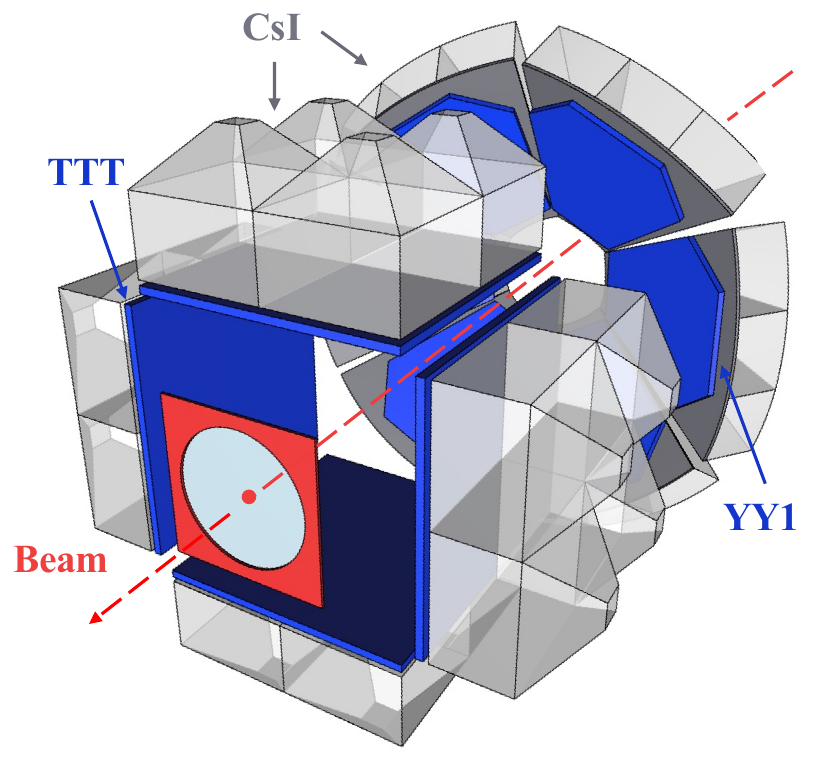}
  \caption{TiNA2 charged particle detector array in backwards configuration. The double-sided (TTT) and single-sided (YY1) silicon strip detectors are colored in blue and the CsI crystals in gray. }
  \label{fig:tina}
\end{figure}

 The TiNA2~\cite{schrock18,mauss21} setup consists of a box of double-sided silicon strip detectors (DSSSDs, Micron type TTT) with $128\times128$ strips and an active area of $97.22 \times 97.22$~mm$^2$ covering the angular range from $103^\circ$ to $154^\circ$ in the laboratory system. Due to the large number of electronic channels the readout was performed using the GET electronics system~\cite{pollacco18}. Positioned behind each DSSSD square are four CsI(Tl) detectors used to detect high-energy charged particles that punch through the silicon, although the protons resulting from the transfer reaction rarely penetrated the 300 $\mu$m thick DSSSDs. At the most backward angles, an array of six trapezoidal single-sided silicon strip detectors (SSDs, Micron type YY1) was installed in a lampshade configuration. These detectors cover laboratory angles up to $170^\circ$ and are also backed with CsI(Tl) crystals which were not required for the present work. The SSDs and the CsI crystals were readout using conventional electronics. A schematic of the setup is shown in Fig.~\ref{fig:tina}. 
In the case of the DSSSDs 95.7 $\%$ of the front and 99.0 $\%$  of the back strips were operational during the experiment, while for the SSDs this number was 88.5 $\%$ of the strips. The overall angular coverage of TiNA2 accounts to around $35\%$ of the total $4\pi$ solid angle. The efficiency and excitation energy resolution of the array were evaluated through detailed Geant4~\cite{agostinelli03} simulations implemented in the NPTool~\cite{matta16} package.

\subsection{Reaction product detection in SHARAQ}
Ejectiles were measured by the SHARAQ spectrometer positioned downstream of the target. For this experiment, the two quadrupole magnets and the first dipole magnet of SHARAQ were employed. The focal plane, S1, was equipped with a pair of SR-PPACs and a 30-fold segmented ionization chamber. The magnetic rigidity of the SHARAQ spectrometer was set to $B\rho = 1.435 (1.367)$~Tm in 2022 (2024).
The beam's large emittance caused it to be defocused in the vertical ($y$) direction at the target, resulting in substantial transmission losses at the target aperture and through SHARAQ. In 2022, only 14.7\% of the \nuc{50}{Ca} ions impinging on the CD$_2$ secondary target were transmitted to the SHARAQ S1 focal plane, in 2024 this transmission was improved to 35.8\%.
Particle identification by mass-to-charge ratio ($A/q$) was achieved using time-of-flight measurements and the position information at the S1 focal plane. The atomic number ($Z$) was identified through the analysis of the energy loss measured in the ionization chamber. The calibration of the energy loss was performed by directing a 44~$A$MeV high-energy beam into the SHARAQ spectrometer. After higher-order aberration corrections, the resolution for the proton number amounted to $\sigma(Z)=0.6$.

\section{Data Analysis}
\subsection{Beam tracking and identification}
\begin{figure}[!h]
  \includegraphics[width=\columnwidth]{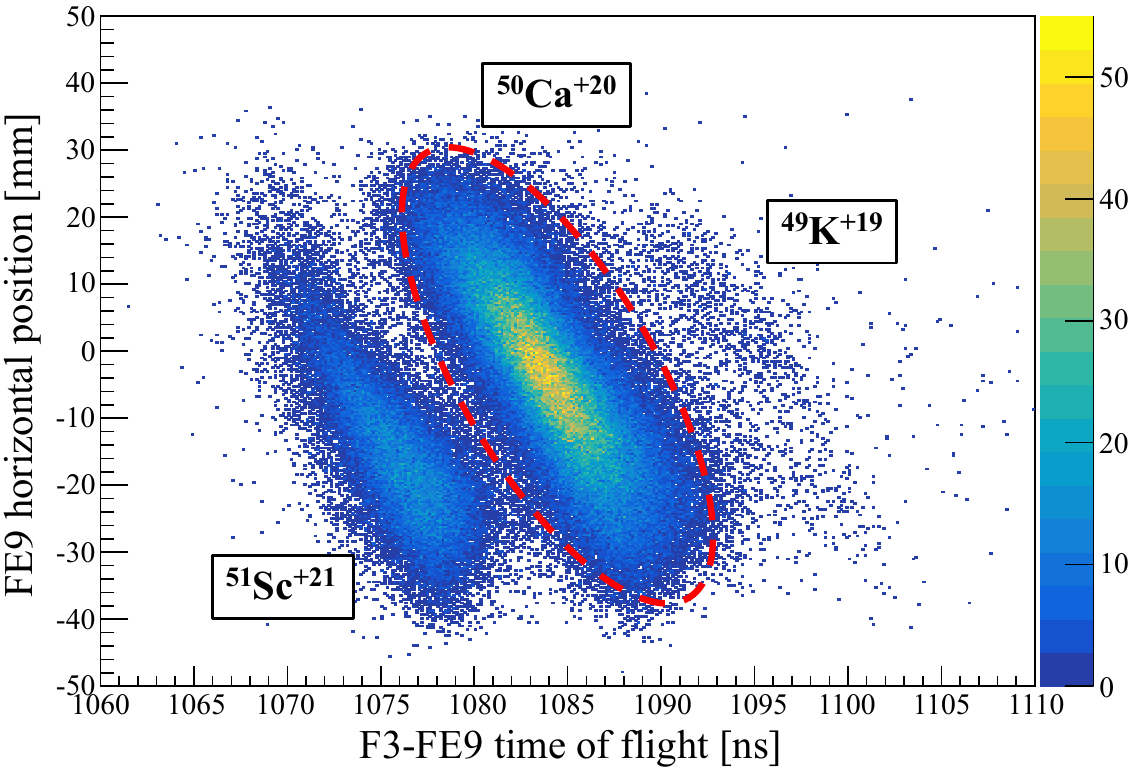}
  \caption{Particle identification in BigRIPS through measurement of the time-of-flight from F3 to FE9 and the position in the FE9 dispersive focal plane.}
  \label{fig:BR_id}
\end{figure}

Particles incident on the secondary target were identified via their time-of-flight between the F3 and FE9 focal planes, as well as their FE9 horizontal position (proportional to the magnetic rigidity $B\rho$). A sample particle identification plot is shown in Fig.~\ref{fig:BR_id}, 
where excellent \nuc{50}{Ca} identification is observed.

\begin{figure}[!h]
  \includegraphics[width=\columnwidth]{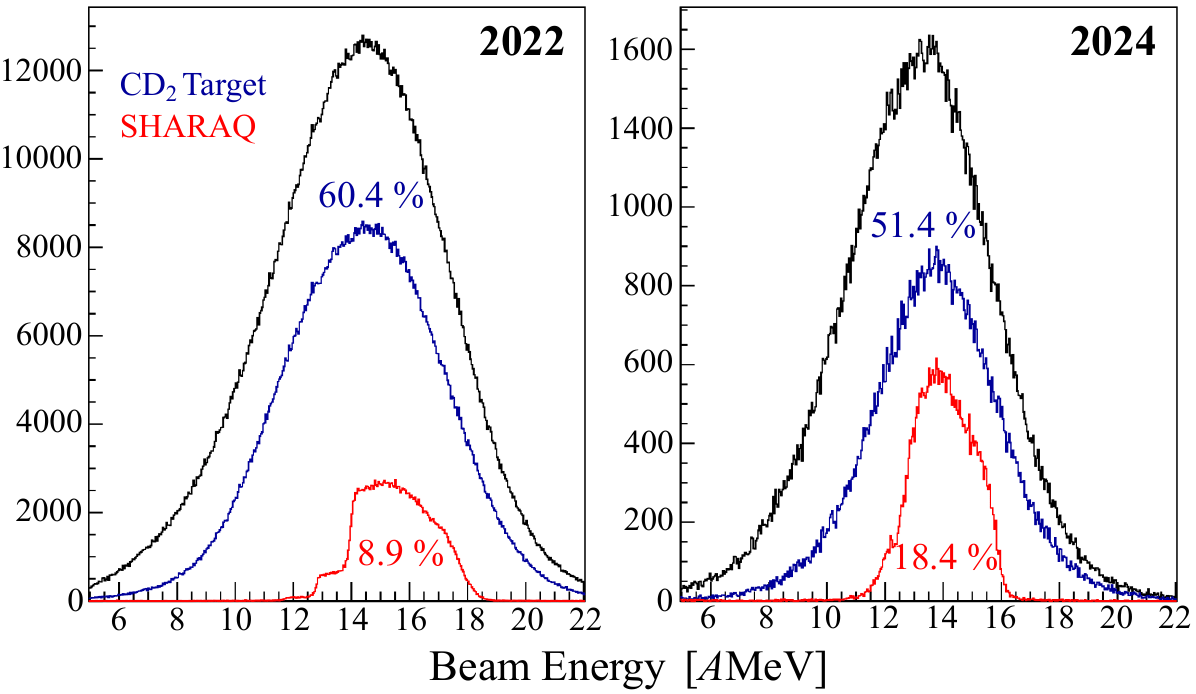}
  \caption{Target midpoint kinetic-energy distribution of all \nuc{50}{Ca} beam ions (black line), those ions reaching the CD$_2$ target (blue), and finally only those which were transmitted to the SHARAQ spectrometer (red). The apparent difference in statistics between the two experiments is mainly due to the choice of a larger trigger downscale factor in 2024.}
  \label{fig:beamenergy}
\end{figure}

The time-of-flight between FE9 and FE12 provided an event-by-event determination of the kinetic energy of the beam, which is extrapolated via ATIMAv1.2~\cite{atima} calculations to the midpoint of the secondary target. The mid-target energy distribution of incoming \nuc{50}{Ca} is shown in Fig.~\ref{fig:beamenergy}, 
where the subsets of ions that hit the secondary target, and eventually reach the SHARAQ spectrometer, are also represented. Although the OEDO focusing of the beam on the CD$_2$ target is superior in 2022 (60.4\% as compared to 51.4\% in 2024), the overall SHARAQ transmission was doubled in 2024 (from 8.9\% to 18.4\%). The transmission losses considerably reduce the width of the beam energy distribution behind the secondary target. 
Since the four-momentum of each incident particle is measured, the scattering angle and kinetic energy of light particles detected in TiNA2 allows the extraction of the excitation energy of the heavy ejectile.

%

\subsection{Recoil detection and missing mass reconstruction}
The silicon detectors of TiNA2 were calibrated using a mixed triple-$\alpha$ source. To identify proton hit candidates, the energies measured in the front and back strips of the DSSSDs were required to agree within 250~keV.  Inter-strip hits and charge-sharing events were recovered by summing the calibrated charge signals from adjacent strips after applying a strip-dependent linear correction factor. The backing CsI detectors were calibrated with high-energy protons coming from the interaction of the beam with the Al target ladder used to hold the secondary target. The energy measured by the CsI detectors was used to detect punch-through events (above $6 - 7$ MeV due to the effective thickness dependence on the incident angle of each scattered proton). In case of punch through, the total proton laboratory energy was calculated from the energy deposition and angle of incidence on the silicon detector. This method was also applied to those events for which based on the kinematics of the transfer reaction punch through would be expected, but due to threshold effects and incomplete angular coverage no CsI energy deposition was recorded. The geometric efficiency of TiNA2 was taken into account in the simulations. To suppress electronic noise that could mimic low-energy proton hits, a threshold of 1.5 MeV was applied to the two detector groups. Coincidence events were selected by requiring beam particles to be detected in the different beamline detectors within a strict 100~ns-wide time gate. 


Since the TiNA2 array is positioned at laboratory angles larger than $90^\circ$, it primarily detects protons originating from the $(d,p)$ transfer reaction. Estimations made using the PACE4 code~\cite{gavron80} demonstrated that in the present analysis the contribution of protons emitted following
fusion-evaporation reactions on the carbon atoms in the CD$_2$ target is negligible. Due to the high beam energy, these protons are forward focused
and furthermore their kinetic energy is below 2~MeV in the angular range covered by TiNA2. Therefore, all events that survive the selection criteria described above were thus considered to originate from the \nuc{50}{Ca}$(d,p)$\nuc{51}{Ca} reaction. To reconstruct the missing mass of the binary reaction, the measured kinetic energy of the protons was corrected for the energy loss in the target material, assuming that the reaction took place at the target midpoint.

\subsection{SHARAQ analysis}
\begin{figure}[!h]
  \includegraphics[width=\columnwidth]{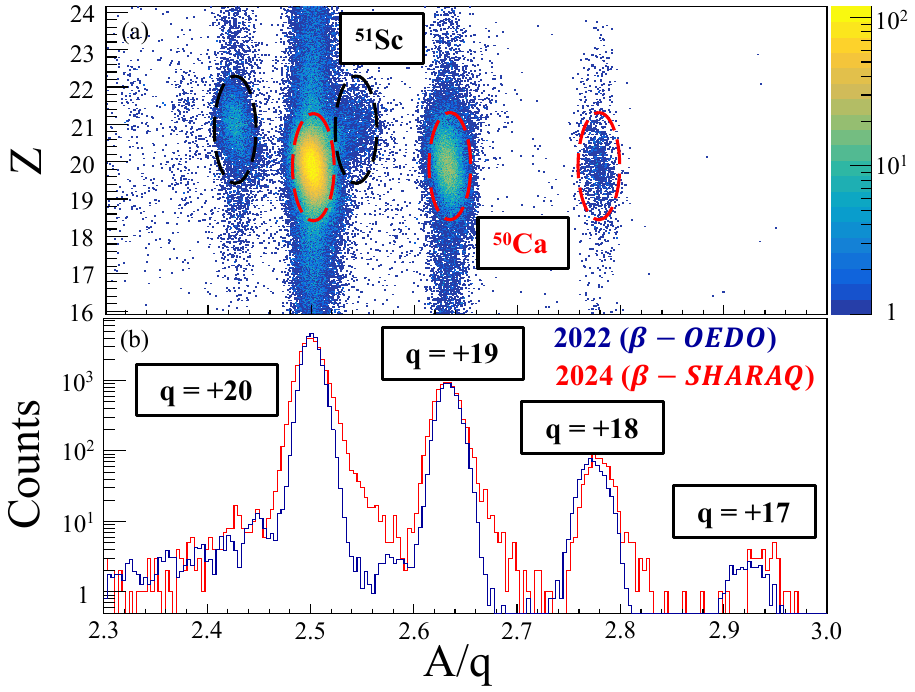}
  \caption{(a) Particle identification of the secondary beam in the SHARAQ spectrometer. (b) Mass-to-charge ratio gated on incoming \nuc{50}{Ca} beam particles detected in the SHARAQ spectrometer for 2022 and 2024. Outgoing \nuc{50}{Ca} in different charge states is identified.}
  \label{fig:aq}
\end{figure}

Reaction products were identified at the SHARAQ S1 focal plane, located downstream of the target chamber. The identification was based on the magnetic rigidity ($B\rho$), velocity (determined from the flight path length and time-of-flight), Bragg peak energy, and range information obtained from the segmented ionization chamber.
The values of $B\rho$ and the flight path through SHARAQ were reconstructed from the particle trajectories recorded by the SR-PPAC detectors at FE12 and S1. The mass-to-charge ratio ($A/q$) of the ions was derived from the $B\rho$ and their velocities $\beta_\text{SHARAQ}$, calculated between the S0 focal plane (extrapolated from FE12 timing and tracking data) and the S1 plane. Additionally, the beam velocity between the FE9 and FE12 focal planes was extrapolated to the S1 focal plane through event-by-event energy-loss calculations for an alternative ($A/q$) determination ($\beta_\text{OEDO}$). The ($A/q$) particle identification plots, in which \nuc{50}{Ca} in different atomic charge states is identified, are shown in Fig.~\ref{fig:aq}. 
Fully stripped, $q=Z=+20$, as well as hydrogen- and helium-like ions with $q=Z-1=+19$ and $q=Z-2=+18$, respectively, are transmitted through SHARAQ. 
While in 2022, the $A/q$ resolution obtained from the extrapolated $\beta_\text{OEDO}$ resulted to be superior, the opposite is true in 2024. This is attributed to differences in the configuration of the D1 dipole magnet between the two experiments.

\section{Results}
\subsection{Missing-mass spectrum}
\begin{figure}[!h]
  \includegraphics[width=\columnwidth]{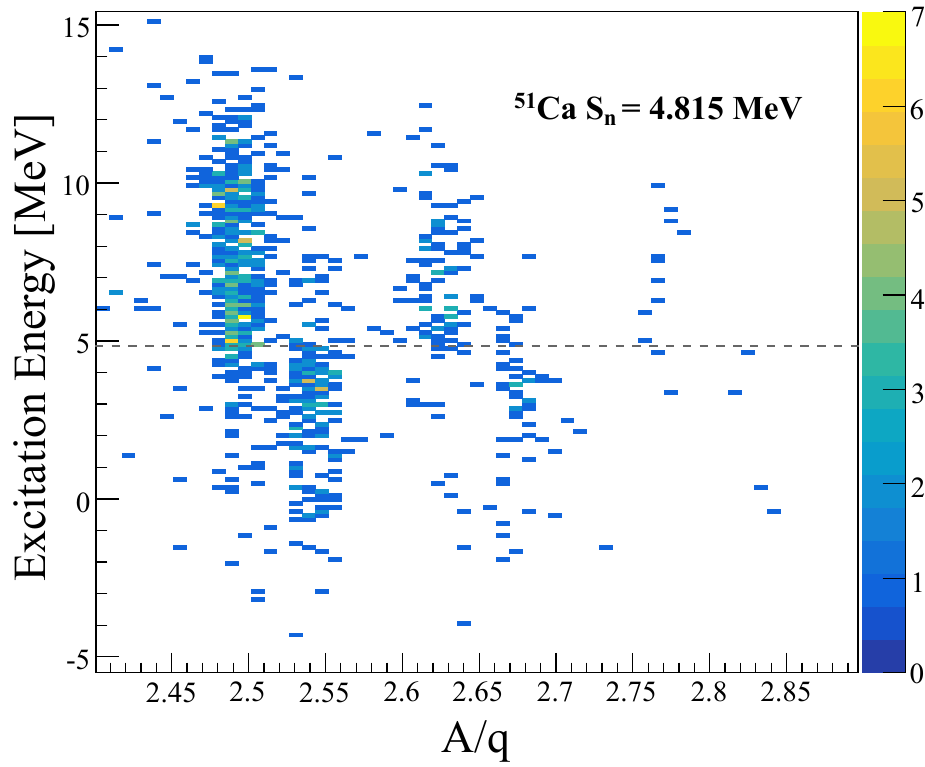}
  \caption{Excitation energy vs $A/q$, A transition in $A/q$ is observed at the neutron separation energy $S_n = 4.815$~MeV~\cite{wang21}. Data from the forward box detectors from 2024 are shown.}
  \label{fig:aqmiss}
\end{figure}

\begin{figure}[!h]
  \includegraphics[width=\columnwidth]{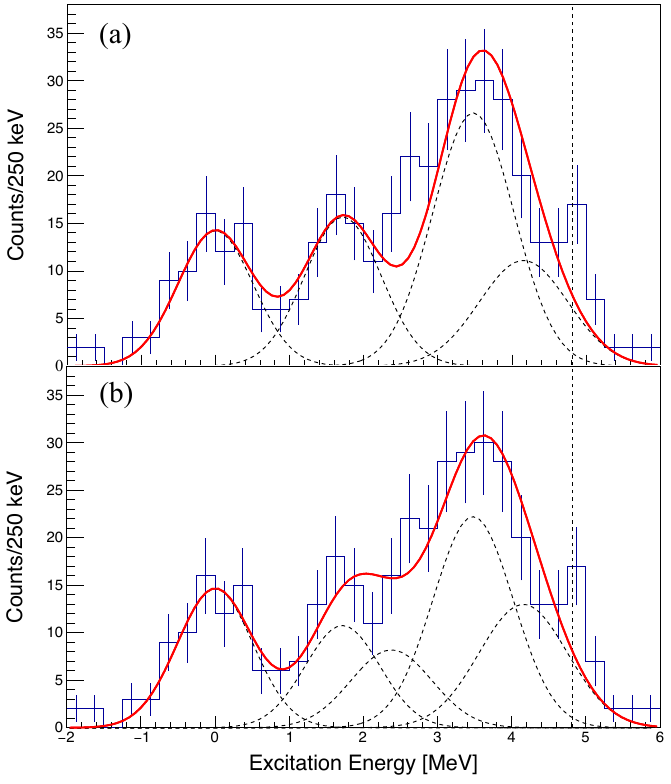}
  \caption{Excitation-energy spectrum of the \nuc{50}{Ca}$(d,p)$\nuc{51}{Ca} transfer reaction adjusted assuming the population of the four states at
  0, 1.718, 3.478, and 4.155~MeV (top, $\chi^2$/NDF = 1.197) and including in addition also the level at 2.378~MeV (bottom, $\chi^2$/NDF = 1.005).}
  \label{fig:spectra}
\end{figure}

After energy calibration, hits in the TiNA2 array were correlated with beam particles identified as shown in Fig.~\ref{fig:BR_id} by selecting coincident events using the timing information recorded. Only events in which the beam ion impacted on the target area were considered for the missing-mass calculation. Additional gates on reaction products, see Fig.~\ref{fig:aq}, in SHARAQ were applied to purify the data.
Fig.~\ref{fig:aqmiss} shows the excitation energy determined from the missing mass reconstruction assuming the binary \nuc{50}{Ca}$(d,p)$\nuc{51}{Ca} reaction versus the mass-to-charge ratio of the reaction products.
Transfer to bound states of \nuc{51}{Ca} results in events characterized by $A/q=2.55$ for fully stripped ions and $A/q=2.68$ for hydrogen-like ions. If the transfer reaction populates states above the neutron separation energy of \nuc{51}{Ca} ($S_n = 4.815$~MeV~\cite{wang21}), neutron emission leads to \nuc{50}{Ca} with $A/q=2.50 (2.63)$ for fully stripped (hydrogen-like) ions. The kinetic energy of protons corresponding to the population of unbound states lies below the detection threshold for a large fraction of the covered angular range, and therefore unbound states are not further analyzed in this work.
The excitation-energy spectrum for bound states, which includes the data from both experiments, is shown in Fig.~\ref{fig:spectra}.
The energy resolution in this spectrum is affected by the energy loss straggling of the beam particles and the protons in the secondary target, the intrinsic energy and position resolution of the TiNA2 detectors, and the uncertainties involved in the position tracking of incident particles onto the target plane. It therefore depends on the excitation energy of the populated state, $E_x$, as well as the polar angle of the proton detection, $\theta_{lab}$. Detailed Geant4 simulations \cite{agostinelli03,matta16}, in which both the TiNA2 array geometry and the secondary targets were implemented, were used to demonstrate that the use of Gaussian distributions to describe the individual
components of the excitation-energy spectrum is a valid approach for the present experiment. To obtain a coherent fit of the excitation energy spectra, the beam tracking resolution on the secondary target was modeled by a two-dimensional Gaussian function with $\sigma \approx$ 5 mm. This in turn allowed the line widths to be parameterized as a function of $E_x$ and $\theta_{lab}$. In a first step, only the population of four states that were predicted to have large spectroscopic overlaps in shell-model calculations (discussed below) were considered, namely the $3/2^-$ ground state, the $(1/2^-)$ state at 1.718 MeV, the $(5/2^-)$ state at 3.478~MeV, and the $(9/2^+)$ state at 4.155~MeV.
Thus, the excitation-energy spectrum was described by four Gaussian distributions with fixed positions and considering the yields as free parameters. The result is shown in Fig.~\ref{fig:spectra}\,(a). Even taking into account the considerable statistical uncertainties, it seems that the shape of the main peak at higher excitation energies is not as well described as the first two peaks in the spectrum. Therefore, an alternative fit was performed, see Fig.~\ref{fig:spectra}\,(b), in which in addition to the four states mentioned above also the first $(5/2^-)$ state at 2.378~MeV, for which a $(1p_{3/2})^2_{J=2}(1p_{1/2})^1$ configuration had been proposed in the past~\cite{rejmund07}, is included.
Since the latter assumption seems to better reproduce the experimental spectrum, in the following discussion the population of five bound states is adopted.

\subsection{Differential cross sections}
\begin{figure}[!h]
  \includegraphics[width=\columnwidth]{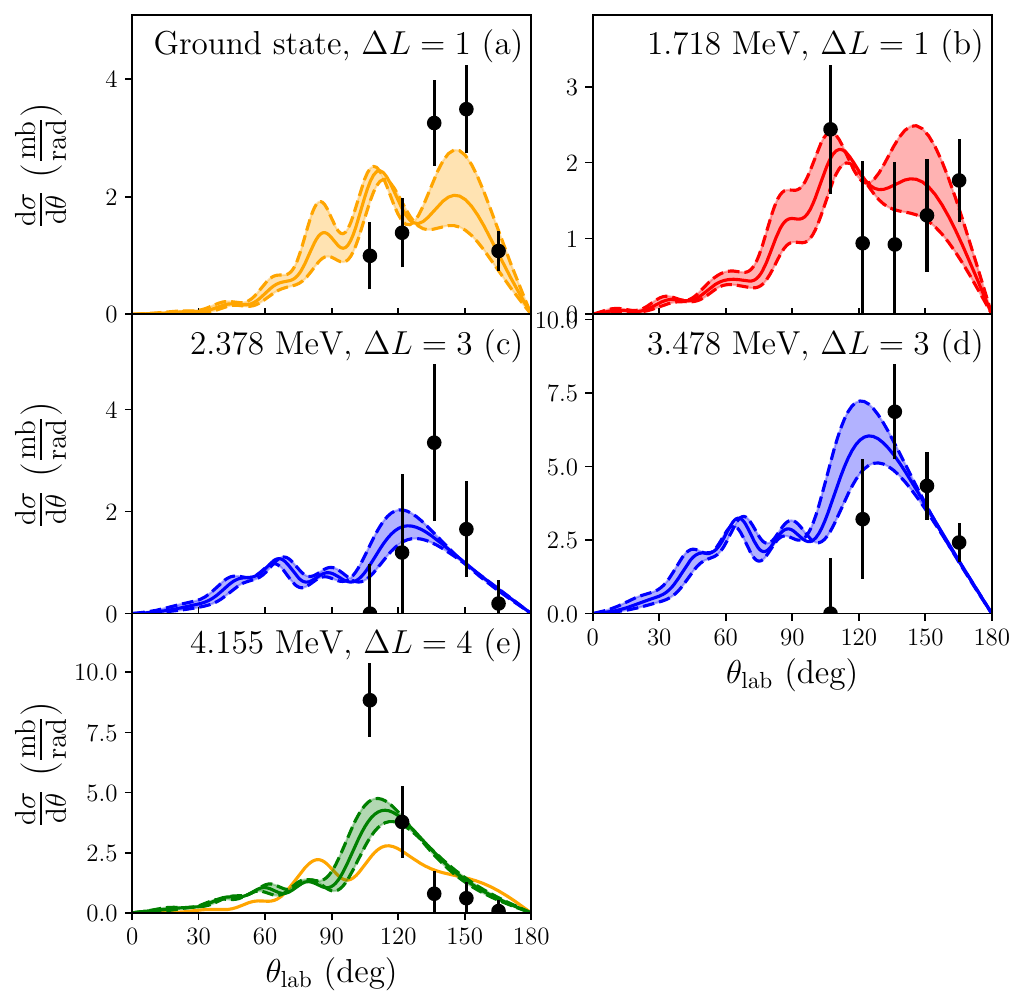}
  \caption{Differential cross sections for the four states populated in \nuc{51}{Ca}. The data in black are overlaid with reaction model calculations based on ADWA. See text for details. For the 4.155~MeV state in (e), a $\Delta L = 1$ transition is shown for comparison.}
  \label{fig:cs}
\end{figure}

In order to derive the differential cross sections, the data was divided into five angular bins. The excitation-energy spectrum for each bin was individually adjusted by Gaussian peaks with fixed positions and widths according to the parametrization obtained based on the simulations. The latter were also used to calculate the solid angle of each bin. Prior to summing the data from the 2022 and 2024 experiments, the measured yields were normalized to the number of incident \nuc{50}{Ca} particles and corrected for trigger efficiency and acquisition live-time.
The differential cross sections extracted assuming the population of five states are shown in Fig.~\ref{fig:cs}.
The error bars include statistical uncertainties from the number of incident particles as well as the protons detected and identified in the TiNA2 detectors. Systematic uncertainties related to the trigger, transmission, and data acquisition deadtime are negligible.

\section{Discussion}
The \nuc{50}{Ca}$(d,p)$\nuc{51}{Ca} transfer reaction was modeled as a single-step process, where the transferred neutron populates an unoccupied valence orbital. By comparing the experimental cross sections for each final state to theoretical calculations, the spectroscopic factor $C^2S$ can be extracted. Due to the relatively high beam energy, the adiabatic distorted wave approximation (ADWA) was chosen. The ADWA model explicitly accounts for the breakup of the loosely bound deuteron. For the incoming deuteron channel, global nucleon-nucleus optical model parameters from Ref.~\cite{koning03}, evaluated at half the beam energy, were used. Calculations were performed using the FRESCO code~\cite{thompson88}.
The calculations for the mean beam energy (a weighted average from the two experiments) as well as for a one-sigma variation of the beam energy are overlaid with the data in Fig.~\ref{fig:cs}.
Of the states observed in the present study, only the ground state of \nuc{51}{Ca} has a firmly established spin and parity assignment~\cite{garciaruiz15}. Based on systematics, the next excited state at 1.718~MeV is expected to have $J^\pi = 1/2^-$. Shell-model calculations predict two $5/2^-$ states, of which the lower-lying one is dominated by a $(1p_{3/2})^2_{J=2}(1p_{1/2})^1$ configuration and therefore is not expected to be strongly populated in the present transfer reaction. For the second $5/2^-$ state, a $0f_{5/2}$ configuration is expected. 
The angular distributions presented in Fig.~\ref{fig:cs}\,(a-d) assume transfer to the valence $1p_{3/2}$, $1p_{1/2}$, and $0f_{5/2}$ orbitals, respectively. Despite the substantial experimental uncertainties, the comparison with the ADWA calculations supports these assignments. For instance, for the 3.478~MeV state, the $\chi^2$ value is 20.1 when assuming transfer to a $5/2^-$ state, whereas the $3/2^-$ hypothesis results in a significantly larger $\chi^2$ of 37.3.
The differential cross section of the $4.155$~MeV state peaks at laboratory scattering angles near 90$^\circ$, suggesting a large angular momentum transfer $\Delta L$. Among the shell-model orbitals relevant to neutron-rich Ca isotopes, the best fit is obtained assuming transfer to the $0g_{9/2}$ orbital. However, due to the limited statistics and angular coverage of the present measurement, a $\Delta L = 1$ transition cannot be completely excluded. For comparison, the assumption of transfer to the $p$ orbitals is also shown in Fig.~\ref{fig:cs}\,(e).

Normalization of the theoretical cross sections to the measured data yields the experimental spectroscopic factors $C^2S$. The extracted spectroscopic factors, along with excitation energies, spin-parity assignments, and theoretical predictions, are summarized in Table~\ref{tab:results}.
\begin{table*}[htb]
  \caption{\label{tab:results} Summary of excitation energies, spin-parity assignments, calculated single-particle cross sections, and spectroscopic factors from experiment and theory. $\sigma_\text{sp}$ is calculated using the optical potentials from Ref.~\cite{koning03} at the mean beam energy of $14.5~A$MeV. The uncertainties in $\sigma_\text{sp}$ reflect the spread of the beam energy within one standard deviation.}
\begin{ruledtabular}
\begin{tabular}{lcrl*{2}{rr}}
  $E$ (MeV) & $J^{\pi}$ & $\sigma_\text{sp}$ (mb) & $C^2S_\text{exp}$ & $E$ (MeV) & $C^2S$ & $E$ (MeV) & $C^2S$ \\
  & & & & \multicolumn{2}{c}{IM-SRG} & \multicolumn{2}{c}{GXPF1Br} \\
\hline
     0 & $3/2^-$ & $12.8^{+3.0}_{-2.3}$ & 0.23(4)  & 0     & 0.51 &     0 & 0.48   \\
 1.718 & $1/2^-$ &  $6.0^{+1.6}_{-1.2}$ & 0.47(14) & 2.185 & 0.92 & 1.554 & 0.87   \\
 2.378 & $5/2^-$ & $20.8^{+1.7}_{-1.5}$ & 0.11(5)  & 2.751 & 0.02 & 2.293 & 0.05   \\
 3.478 & $5/2^-$ & $22.9^{+2.4}_{-2.1}$ & 0.35(7)  & 4.144 & 0.95 & 3.594 & 0.84   \\
 4.155 & $9/2^+$ & $29.0^{+2.3}_{-1.9}$ & 0.15(3)  & 7.941 & 0.80 & 3.944 & 0.30   \\
\end{tabular}
\end{ruledtabular}
\end{table*}
The uncertainties in $C^2S$ include both statistical uncertainties from the experimental data and uncertainties due to the beam energy spread; these contributions were added in quadrature. 
In addition to these, the extracted spectroscopic factors are subject to systematic theoretical uncertainties arising from the choice of reaction model, optical model parameters, and the bound-state potential used for the transferred neutron. Based on comparisons using different parameterizations available in the literature, this systematic uncertainty is estimated to be approximately 10\%.
When the $5/2^-$ state at 2.378~MeV is excluded from the fit, see Fig.~\ref{fig:spectra}\,(a), the extracted spectroscopic factors vary by less than the statistical uncertainty. 

The ground state of \nuc{50}{Ca} is expected to have a predominantly $\nu(1p_{3/2})^2$ configuration, as predicted by shell-model calculations and supported by experimental cross sections from single-neutron removal reactions on a \nuc{50}{Ca} beam~\cite{crawford17}. For the present $(d,p)$ transfer reaction, within an independent-particle model, one would therefore expect a spectroscopic factor of $C^2S = 0.5$ for the half-filled $1p_{3/2}$ orbital, and $C^2S = 1$ for the unoccupied $1p_{1/2}$ and $0f_{5/2}$ orbitals.
Taking into account the known quenching of single-particle strength in direct reactions~\cite{aumann21} relative to independent-particle or configuration-interaction shell-model calculations (with a reduction factor $R \sim 0.6$), the spectroscopic factors extracted for the ground and first excited states of \nuc{51}{Ca} are in good agreement with expectations and leave no room for significant single-particle strength in higher-lying $3/2^-$ and $1/2^-$ states. On the other hand, the spectroscopic factor of $C^2S_\text{exp} = 0.35(7)$ extracted for the 3.478~MeV state is somewhat smaller than expected for the unoccupied $0f_{5/2}$ orbital. 
For the $(5/2^-_1)$ state, which is expected to be a $3qp$ state, a spectroscopic factor of $C^2S_\text{exp} = 0.11(5)$ was obtained. The sum of the spectroscopic strength for both $(5/2^-)$ states amounts to $0.46(9)$.
Assuming the 4.155~MeV state to be of $0g_{9/2}$ nature, the spectroscopic factor is 0.15(3). In contrast, $\Delta L =1$ transfer would result in spectroscopic factors of 0.98 for the $1p_{1/2}$ orbital and 0.49 for the $1p_{3/2}$ orbital, effectively exhausting the available single-particle strength. This further supports the assignment of $J^\pi = 9/2^+$ for the 4.155~MeV state.

\begin{figure}[htb]
  \includegraphics[width=\columnwidth]{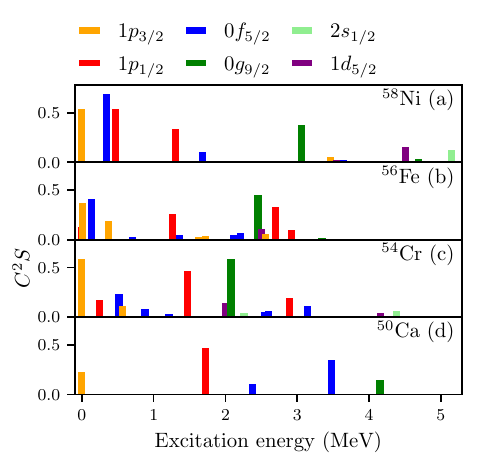}
  \caption{
    Spectroscopic strengths for neutron addition for the $N=30$ isotones \nuc{58}{Ni}, \nuc{56}{Fe}, \nuc{54}{Cr}, and \nuc{50}{Ca}, as a function of excitation energy. 
 Data for neutron spectroscopic factors are taken from  $(d,p)$ transfer reactions~\cite{iwamoto94,decken73,sengupta71,thomson74,macgregor72}.}
  \label{fig:specfac}
\end{figure}

The spectroscopic strength distribution determined in the present work is compared to other $N=30$ isotones in Fig.~\ref{fig:specfac}.
Compared to its isotones, the level density in \nuc{51}{Ca} is significantly lower. While a number of levels are experimentally established in the energy range around 4~MeV~\cite{perrot06,rejmund07,fornal08} most of the levels possess a complex structure with very small spectroscopic factors for the neutron transfer reactions.
The systematics show that the single-particle strength for the $1p_{3/2}$ is concentrated in the ground state in all $N=31$ isotones. A clear trend is observed for the neutron $0f_{5/2}$ orbital as one moves from the stable doubly-magic nucleus \nuc{58}{Ni} to the neutron-rich \nuc{50}{Ca}: the centroid of the single-particle strength shifts to higher excitation energies, and the strength becomes increasingly fragmented in \nuc{54}{Cr}. No experimental data are currently available for \nuc{52}{Ti}.  In the semi-magic nucleus \nuc{50}{Ca}, the fragmentation is reduced again. In the present study, a large fraction of the expected spectroscopic strength for the $0f_{5/2}$ orbital is fragmented in two states in \nuc{51}{Ca}.
The $1p_{1/2}$ strength is located in two states for $Z=28$ and 30 while more fragmentation is observed for the deformed \nuc{54}{Cr}. Regarding the positive parity $0g_{9/2}$ orbital, low-lying states with significant strength, $C^2S\sim 0.5$ are observed in \nuc{59}{Ni}, \nuc{57}{Fe}, and \nuc{55}{Cr}. The state with the major fraction of the single-particle strength is slightly lowered in energy when removing protons from the magic Ni. While no experimental data exists for \nuc{53}{Ti}, this trend is broken in \nuc{51}{Ca}. The first $9/2^+$ state is located at a higher excitation energy, and its population in the $(d,p)$ transfer reaction is significantly reduced compared to the heavier isotones. This situation is different in the $N=29$ isotopes, where the spectroscopic factor for the first $9/2^+$ state is linearly reduced from $C^2S = 0.74$ in \nuc{55}{Fe}~\cite{fulmer63} to 0.14 in \nuc{49}{Ca}~\cite{uozumi94}. At the same time, the excitation energy remains almost constant at 3.7~MeV and increases only at $Z=20$ to above 4~MeV. 
It should be noted, that the spectroscopic strength determined in heavy-ion induced pickup reactions on \nuc{48}{Ca} was determined to be larger with $C^2S=0.27$~\cite{gade16} compared to the $(d,p)$ transfer reactions.
The $9/2^+$ states in Ca isotopes have also been discussed as proton dominated octupole deformed~\cite{riley16}. This could explain the reduced neutron spectroscopic strength and would suggest that the major fraction of the $\nu 0g_{9/2}$ orbital lies above the separation energy. However, in \nuc{49}{Ca} the second $9/2^+$ state carries only 2\% of the full $L=4$ strength~\cite{uozumi94}.

\begin{figure}[!h]
  \includegraphics[width=\columnwidth]{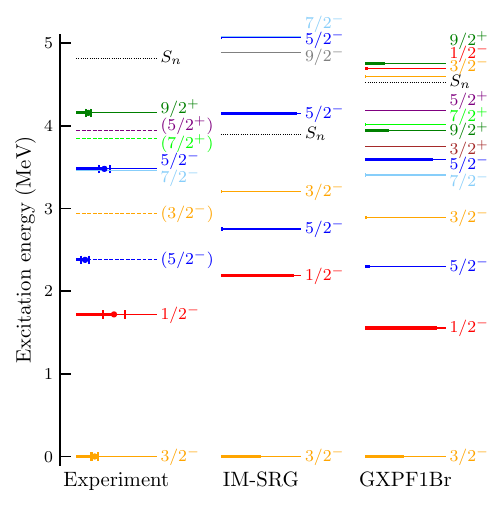}
  \caption{
    Spectroscopic strengths for the \nuc{50}{Ca}$(d,p)$\nuc{51}{Ca} transfer reaction. States are labeled with their total angular momentum $J^\pi$. The length of the bars indicates the magnitude of the spectroscopic factors.}
  \label{fig:specfac_compare}
\end{figure}

The experimental results are compared to shell-model calculations in Table~\ref{tab:results} and Fig.~\ref{fig:specfac_compare}.
The GXPF1Br calculations make use of an effective interaction, based on the GXPF1B~\cite{honma08} interactions, extended to the full $sd-fp-sdg$ model space using the USD-GXPF1B-VMU effective interactions~\cite{utsuno12}. This interaction has been successfully used to describe neutron-rich Ca isotopes~\cite{steppenbeck13}. The agreement with the experimental data is remarkable. Considering the spectroscopic factors, and taking into account the usual reduction observed in experiments, both the neutron single-particle like states, as well as states of more complex nature, including positive parity states, are very well described. These calculations predict a second $9/2^+$ state at 4.748~MeV with substantial spectroscopic strength, $C^2S=0.25$, which is not within the acceptance of the present experiment and could explain the apparent reduction of the $0g_{9/2}$ strength. In the GXPF1Br shell-model calculations, the major fraction of the wave function of the $5/2^-_1$ state in \nuc{51}{Ca} corresponds to coupling a neutron in the $1p_{1/2}$ orbital to the first $2^+$ excited state in the \nuc{50}{Ca} core. For the $9/2^+$  state, the calculations indicate that its structure is a mix of single-particle and built on coupling to the $5^-_1$ state in \nuc{50}{Ca}, rather than the octupole $3^-_1$  excitation as discussed in Ref.~\cite{riley16}. These results underscore the importance of complex core excitations in shaping the level structure of \nuc{51}{Ca}, particularly for states at higher excitation energies.

Valence-space in-medium similarity renormalization group (VS-IMSRG) calculations~\cite{hergert16,stroberg19} employed the 1.8/2.0 (EM) interaction~\cite{hebeler11} within the neutron $\{1p_{1/2}$, $1p_{3/2}$, $0f_{5/2}$, $0g_{9/2}$, $1d_{5/2}$, $2s_{1/2}\}$ valence space above the \nuc{48}{Ca} core.
To obtain the effective Hamiltonian, the Hamiltonian originally expressed within 13 major-shell harmonic-oscillator space was transformed by the approximate two-body level unitary transformation using the modified generator introduced in Ref.~\cite{Miyagi2020}.
The excitation energy of states is slightly over estimated, but the spectroscopic strength is in agreement with the experiment and other theoretical approaches. Only the energy of the $9/2^+$ state is severely overestimated (7.9~MeV) and the calculations predict this state to have strong single-particle character ($C^2S=0.80$) in contrast to the experimental results. 


\section{Conclusion}
In summary, the neutron-rich nucleus \nuc{51}{Ca} was investigated using the \nuc{50}{Ca}$(d,p)$ transfer reaction in inverse kinematics using an
energy-degraded beam at OEDO/RIBF. Several states were populated and reconstructed via missing mass spectroscopy, and their angular distributions were used to extract differential cross sections and spectroscopic factors. The results support the spin-parity assignments of $3/2^-$ to the ground state and of $1/2^-$, $5/2^-$, and $9/2^+$ to the excited states at 1.718, 2.378, 3.478, and 4.155~MeV, respectively, providing new constraints on the single-particle structure of neutron-rich calcium isotopes. The experimental spectroscopic factors are compared with shell-model calculations using the GXPF1Br interactions as well as with valence-space IM-SRG results. The overall agreement confirms the dominant single-particle character of the low-lying negative-parity states. These findings provide valuable benchmarks for nuclear structure models and highlight the evolving shell structure in neutron-rich calcium isotopes.

\acknowledgments
We are grateful for the support of the RIKEN Nishina Center accelerator staff for the reliable delivery of the \nuc{70}{Zn} primary beam. 
This work has been supported by the Deutsche Forschungsgemeinschaft (DFG, German Research Foundation) - Project-ID 279384907 - SFB 1245,
by MCIN/AEI/10.13039/501100011033, Spain, with grant PID2023-150056NB-C42 and PID2020-118265GB-C41, by JST ERATO Grant No. JPMJER2304, Japan, by JSPS KAKENHI Grant Numbers JP19H01914, JP24H00239, and by the Slovenian Research and Innovation Agency under grants No. P1-0102 and I0-E005. This work was also partly supported by the Institute for Basic Science (IBS) funded by the Ministry of Science and ICT, Korea (Grants No. IBS-R031-D1, IBS-R031-Y1, IBS-R031-Y2).
For the VS-IMSRG calculations, the \texttt{NuHamil}~\cite{NuHamil}, \texttt{imsrg++}~\cite{imsrg++}, and \texttt{KSHELL}~\cite{Shimizu2019} codes were used to generate chiral EFT matrix elements, to perform the valence-space decoupling, and to solve the valence-space problems, respectively. The CD$_2$ targets were kindly provided by INFN-LNS courtesy of Prof. Silvio Cherubini (Universit\`a degli Studi di Catania and INFN, Laboratori Nazionali del Sud, Catania, Italy).
\bibliography{draft}

\end{document}